\documentclass[journal]{IEEEtran}
\usepackage{amsmath,amssymb}
\usepackage{mathtools}
\usepackage{amsfonts}
\usepackage{amsbsy}
\usepackage{graphics}
\usepackage{subfigure}
\usepackage{graphicx}
\usepackage{lipsum} 
\usepackage{booktabs}
\usepackage{float}

\usepackage[countmax]{subfloat}
\usepackage{epstopdf}
\usepackage{chngcntr}
\usepackage{wrapfig}
\renewcommand{\thispagestyle}[1]{} 
 
\usepackage[acronym]{glossaries}
\usepackage{tabularx}
\usepackage{cite,url}
\usepackage{hyperref}
\usepackage{multirow}
\usepackage{longtable}
\usepackage[most]{tcolorbox}
\usepackage{amsmath,amssymb,amsfonts}
\usepackage{bm}
\usepackage{bbm}
\usepackage{graphicx}
\usepackage{cite}
\usepackage{epstopdf}
\usepackage{balance}
\usepackage{gensymb}
\usepackage{lipsum}
\usepackage{float}
\usepackage{epstopdf}
\usepackage[mathcal]{eucal}
\usepackage{subfiles}
\usepackage[T1]{fontenc}
\usepackage{siunitx}
\usepackage{tabulary}

\usepackage{epsfig,graphics,graphicx,amssymb,amstext,amsmath,algorithm,algorithmic,wrapfig,multirow}
\usepackage{array}
\usepackage{adjustbox}
\usepackage{xcolor}
\usepackage[font=small,labelfont=bf,labelsep=colon]{caption}
\usepackage{tabularx}
\usepackage{bm}
\usepackage{enumerate}
\usepackage{tikz}
\usepackage{pgfplots}
\usepackage{epstopdf}
\usetikzlibrary{shapes,arrows}
\usepackage[utf8]{inputenc}
\usepackage{mathtools}
\usepackage[utf8]{inputenc}
\usepackage[algo2e,ruled, vlined, linesnumbered]{algorithm2e}
\usepackage{tikz}
\usepackage{multirow}
\usetikzlibrary{chains,arrows,calc,positioning}
\usepackage{amssymb}
\usepackage{pifont}
\usepackage{caption}
\usepackage{subcaption}

\def\BibTeX{{\rm B\kern-.05em{\sc i\kern-.025em b}\kern-.08em
		T\kern-.1667em\lower.7ex\hbox{E}\kern-.125emX}}
\usepackage{soul}

\usepackage{breqn}
\usepackage{booktabs}
\usepackage{multirow}
\usepackage{enumitem}
\usepackage{tabulary}
\usepackage[normalem]{ulem}

\makeatletter

 \let\oldforeign@language\foreign@language
 \DeclareRobustCommand{\foreign@language}[1]{%
   \lowercase{\oldforeign@language{#1}}}

\usepackage{dblfloatfix}
\usepackage{makecell}

\makeatother

\IEEEoverridecommandlockouts
\begin{document}
\title{Optimal Linear Precoding Under Realistic Satellite Communications Scenarios}
\author{{Geoffrey~Eappen}, \IEEEmembership{Member,~IEEE}, 
 {Jorge~Luis~Gonzalez}, \IEEEmembership{Member,~IEEE},
{Vibhum Singh}, \IEEEmembership{Member,~IEEE},
{Rakesh~ Palisetty}, \IEEEmembership{Member,~IEEE}, {Alireza~Haqiqtnejad}, \IEEEmembership{Member,~IEEE}, {Liz~Martinez~Marrero}, \IEEEmembership{Member,~IEEE}, {Jevgenij~Krivochiza}, \IEEEmembership{Member,~IEEE}, {Jorge~Querol}, \IEEEmembership{Member,~IEEE}, {Nicola~ Maturo}, \IEEEmembership{Member,~IEEE}, {Juan~Carlos~Merlano~Duncan}, \IEEEmembership{Senior~Member,~IEEE}, {Eva~Lagunas}, \IEEEmembership{Senior Member,~IEEE}, {Stefano~Andrenacci}, and {Symeon~Chatzinotas}, \IEEEmembership{Fellow,~IEEE}

\thanks{G.  Eappen, J. L. Gonzalez, V. Singh,  L. M. Marrero,  J.  Querol, J. C. M. Duncan, E. Lagunas,  and S. Chatzinotas are with the Interdisciplinary Centre for Security Reliability and Trust, University of Luxembourg, 1855 Luxembourg City, Luxembourg (e-mails: \{geoffrey.eappen, jorge.gonzalez,
vibhum.singh, liz.martinez-marrero, jorge.querol, juan.duncan, eva.lagunas,  symeon.chatzinotas\}@uni.lu).}
\thanks{Rakesh Palisetty is with the Department of Electrical Engineering, Shiv Nadar Institution of Eminence Deemed to be University, Delhi NCR 201314, India (e-mail:rakesh.palisetty@snu.edu.in).}
\thanks{A. Haqiqtnejad is with OQ Technology, 1 rue de la Poudrerie, 3364, Leudelange, Luxembourg (e-mail:alireza.haqiqatnejad@oqtec.com ).}
\thanks{J. Krivochiza and S. Andrenacci  are with SES S.A. Ch\^{a}teau de Betzdorf, Betzdorf L-6815. Luxembourg (\{e-mails: {Jevgenij.Krivochiza, Stefano.Andrenacci\}@ses.com).}}
\thanks{N. Maturo is with European space Agency-ESA, Leiden, South Holland, Netherlands (e-mail:nicola.maturo@esa.int ).}
\thanks{Corresponding author: Vibhum Singh (email: vibhum.singh@uni.lu).}
\thanks{This work was fully supported by European Space Agency under the project number 4000122451/18/NL/NR ``Live Satellite Demonstration of Advanced Interference Management Techniques (LiveSatPreDem)'' and SES S.A. (Opinions, interpretations, recommendations, and conclusions presented in this paper are those of the authors and are not necessarily endorsed by the European Space Agency or SES). This work was supported in parts by the Luxembourg National Research Fund (FNR), through the CORE Project (ARMMONY): ``\textit{Ground-based distributed beamforming harmonization for the integration of satellite and Terrestrial networks}'', under Grant FNR16352790.}}
\maketitle


\begin{abstract}
In this paper, optimal linear precoding for the multibeam geostationary earth orbit (GEO) satellite with the multi-user (MU) multiple-input-multiple-output (MIMO) downlink scenario is addressed. Multiple-user interference is one of the major issues faced by the satellites serving the multiple users operating at the common time-frequency resource block in the downlink channel. To mitigate this issue, the optimal linear precoders are implemented at the gateways (GWs). The precoding computation is performed by utilizing the channel state information obtained at user terminals (UTs). The optimal linear precoders are derived considering beamformer update and power control with an iterative per-antenna power optimization algorithm with a limited required number of iterations. The efficacy of the proposed algorithm is validated using the In-Lab experiment for 16$\times$16 precoding with multi-beam satellite for transmitting and receiving the precoded data with digital video broadcasting satellite-second generation extension (DVB-S2X)  standard for the GW and the UTs. The software defined radio platforms are employed for emulating the GWs, UTs, and satellite links. The validation is supported by comparing the proposed optimal linear precoder with full frequency reuse (FFR), and minimum mean square error (MMSE) schemes. The experimental results demonstrate that with the optimal linear precoders it is possible to successfully cancel the inter-user interference in the simulated satellite FFR link. Thus, optimal linear precoding brings gains in terms of enhanced signal-to-noise-and-interference ratio,  and increased system throughput and spectral efficiency.
\end{abstract}

\section{Introduction} 

\IEEEPARstart{D}{uring} last decades, the multiple-input-multiple-output (MIMO) communications concept has had a surge in popularity in academia and industry. The MIMO concept, and in particular the massive-MIMO (M-MIMO) concept has been the most salient characteristic of the latest wireless communication standards including the fifth-generation (5G) of mobile radio systems, Wi-Fi, and even power line communications. However, this strong impulse has not been seen yet in the field of satellite communications (SATCOMs). To meet the high user demands, most SATCOM services focus on high throughput satellites (HTSs) with multiple spot beams for enhancing spectral efficiency (SE) and high data rate connectivity. Primarily, in the multiple spot beams, full frequency reuse (FFR) is employed, leading to high SE \cite{minoli}. The major challenge associated with the multiple spot beam is the interference between the adjacent beams owing to the presence of the side lobes in the beam radiation pattern on a particular coverage area \cite{joroughi2018robust}. 

Recent years have witnessed a strong impulse in adopting M-MIMO in wireless networks \cite{ETSITR102376-2}, \cite{1391204}. Despite the wide literature related to M-MIMO for terrestrial networks, much less attention has been devoted to its possible exploitation in the forward link of SATCOM systems. As such, the 5G mobile radio communication systems aim to deliver seamless integration and enhanced flexibility across various telecommunication networks. Traditionally, terrestrial and satellite systems have developed separately, leading to significant technological differences between these networks. However, the vision of 5G heterogeneous networks includes SATCOM to increase the capacity of 5G networks in terms of better coverage, reliability, availability, and scalability. The coexistence of satellites and the 5G networks require interference-free links. The role of precoding schemes and power allocation is critical in guaranteeing interference-free communication with enhanced quality. Therefore, it is important to define standards for incorporating 5G with SATCOM. The 5GPPP research initiative, co-funded by the European Commission, aims to establish new unified standards for 5G networks \cite{ETSITR102376-2}. Within this initiative, the METIS 2020 project focuses on laying the groundwork for next-generation mobile and wireless communication systems, targeting 2020 and beyond \cite{1391204}. These standards will enable the seamless integration of mobile cellular communications and satellite systems into a unified service.

\begin{figure}[t!]
\centering
\includegraphics[height=1.7in, width=3in]{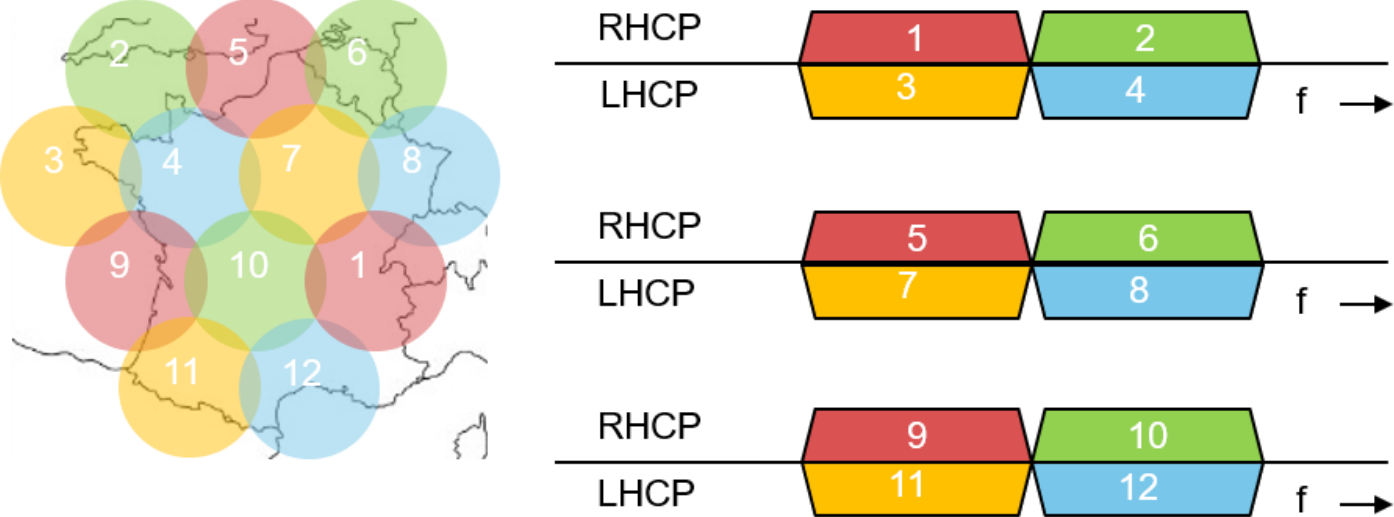}
\caption{Four-color FR (4FR).}
\label{fig:4cr}\vspace{-0.2in}
\end{figure}

Incorporating modern SATCOM systems into 5G networks offers various use cases, such as extending coverage for traditional terrestrial cells, enabling caching through multicast/broadcast data transmission, and providing off-load backhauling for unicast user traffic \cite{5585631}, \cite{opt_lin_prec}. The digital video broadcasting satellite-second generation extension (DVB-S2X) \cite{morello2016dvb} was developed to complement new scenarios for flexible SATCOM integration into 5G and beyond networks. MIMO precoding techniques are based on the closed-loop approach by employing the retrieved channel state information (CSI) from the user terminals (UTs), requiring a feedback channel from UT back to the gateway (GW).  Due to the time-varying channel, the GW has only access to a delayed version of CSI, which can eventually limit the overall system performance \cite{9384465,haqiqatnejad2018robust,demonstrator_nicola2019,tato2018link}.  However, in contrast to general multiuser (MU) MIMO terrestrial systems, the CSI degradation in multibeam mobile applications has a very limited impact on typical fading channel and system assumptions. Under realistic conditions, the numerical results of \cite{haqiqatnejad2018robust} demonstrate that precoding can offer an attractive gain in the system throughput compared with the conservative frequency reuse (FR) allocations. Fig. \ref{fig:4cr} depicts the four-color FR scheme.  MIMO precoding techniques, which are generally defined as closed-form methods or approximate methods, are very quick and less complex. But, these methods provide suboptimal solutions and have to be solved by time-consuming iterative convex optimization (CVX) or non-negative least squares (NNLS) solving methods that must fit into a relevant time frame \cite{clerckx2013mimo}. Recent research is advanced on the reduction of the processing times to meet channel requirements \cite{tato2018link}, \cite{krivochiza2019fpga}. 

Zero-forcing (ZF) precoding was practically demonstrated in \cite{duncan2018hardware, krivochiza2018computationally, merlano2017computationally} using real-time signal processing and transmission. Academic research shows that precoding techniques in SATCOM potentially allow more efficient spectral utilization and substantially higher service availability \cite{mazzali2018enhancing, jiang2016low, lucciardi2017non}. To enable the efficient utilization of satellite transponders, multiple carriers have to be relayed through a single high power amplifier (HPA). However, the non-linear nature of the HPA results in adjacent channel interference and increased peak-to-average power ratio (PAPR), which limits the expected performance gains \cite{chatzinotas2015cooperative}. The symbol-level precoding design proposed in \cite{spano2016per} allows controlling the instantaneous per-antenna transmit power, thus, leading to a reduction of the power peaks, which are detrimental to the aforementioned non-linearity problem. It should be mentioned that this is not possible in the channel-level approach, where the precoder is designed for an entire codeword, including several symbols.  Hence, the transmitted power can be controlled only on an average basis, not on a symbol-by-symbol basis. In the context of nonlinear channels, it is also worth mentioning that more advanced symbol level precoding schemes \cite{spano2016per, spano2017spatial, krivochiza2021end, monzon2022non} which aim at reducing the PAPR of the transmitted waveform, considerably improve their robustness. 

Inspired by the aforementioned discussion, in this paper, we consider an optimal minimum mean square error (MMSE) precoding scheme while employing a proper iterative per-antenna power optimization algorithm with a limited required number of iterations. This paper advances the field by designing, implementing, and experimentally evaluating the performance of MMSE algorithms with per-antenna optimization using a software-defined radio, specifically the universal software radio peripheral (USRP) platform. The study involves constructing a practical communication system, which includes a transmitter, a channel emulator, and a receiver. 

In particular, we focus on creating a hardware demonstrator for a closed-loop 16$\times$16 precoded SATCOM system. We develop the multi-beam DVB-S2X compliant GW, the satellite MIMO channel emulator, and the associated UTs. Further, we validate the design requirements with appropriate software and hardware resources. Notably, the DVB-S2X physical layer is implemented using commercial SDR platforms. The data interface for this DVB-S2X gateway adheres to the “Mode Adaptation input interface,” delivering complete bundled frame packets alongside the corresponding MODCOD configuration \cite{ETSITR102376-1}. All functional components of the DVB-S2X GW—such as scrambling, encoding, framing, modulation, and pulse-shaping—along with the precoding algorithm operate in real-time on the SDR. Utilizing SDR technology allows for rapid prototyping and deployment of the precoded transmission via the optimal MMSE precoding scheme in a realistic setting, as opposed to relying solely on numerical simulations. In summary, our main contributions in this paper can be pointed out as follows:

\begin{itemize}
    \item We explore an optimal MMSE precoding scheme integrated with an efficient iterative per-antenna power optimization algorithm. This algorithm is designed to converge within a limited number of iterations, ensuring computational efficiency.
    \item The performance of the MMSE precoding scheme combined with per-antenna optimization algorithms is further experimentally evaluated using a Software-Defined Radio (SDR) platform, specifically the USRP.
    \item To assess the practical performance of our proposed solutions, we develop a comprehensive communication system. This includes a transmitter, a channel emulator, and a receiver, providing a realistic environment for testing.
    \item A hardware demonstrator for a closed-loop $16\times 16$ precoded satellite SATCOM system is implemented, detailing the design and functionality of a multi-beam DVB-S2X compliant gateway, a satellite MIMO channel emulator, and a set of UTs.
    \item The system design requirements are validated using practical software and hardware resources. This includes demonstrating the real-time execution of DVB-S2X physical layer components such as scrambling, encoding, framing, modulation, and pulse-shaping, all implemented using SDR techniques.
    \item We utilize the ``Mode Adaptation input interface" for the data format of the DVB-S2X gateway, guaranteeing that complete bundled frame packets are delivered along with the appropriate MODCOD configuration.
    \item By leveraging SDR technology, we achieve rapid deployment of the prototype, enabling us to test the optimal MMSE precoding scheme in a realistic environment, moving beyond mere numerical simulations.
\end{itemize}


The rest of this paper is organized as follows. In Section~\ref{sec:syst_model}, we describe the system model. Precoding implementation is presented in Section~\ref{sec:precoding_implementation}.
Section~\ref{sec:experimental} provides details on the experimental validation of the proposed methods, followed by concluding remarks in Section \ref{sec:conclusions}.

\textit{Notations}: The upper-case and lower-case bold-faced letters are used to denote matrices and column vectors, respectively. The superscripts $(\cdot)^{\textmd{H}}$, $(\cdot)^{\dag}$,  and $(\cdot)^{-1}$ represent the Hermitian,  transpose, and inverse operations in the matrix, respectively. Further, $|| \cdot ||$ denotes an absolute magnitude of a complex value, whereas $\textbf{I}$ being the Identity matrix.

\section{System Description} 
\label{sec:syst_model}

\begin{figure}[t!]
\centering
\includegraphics[height=2.2in, width=3in]{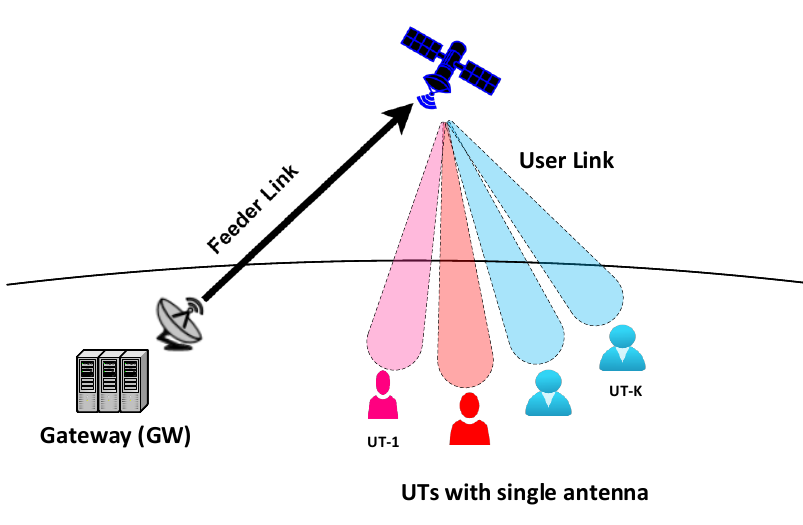}
\caption{System model concept diagram.}
\label{fig:sysmo}\vspace{-0.2in}
\end{figure}

We consider a precoding-enabled geostationary earth orbit (GEO) satellite system consisting of a GW, a transponder capable of generating $N$ beams, and $K (K \leq N)$ single-antenna UTs as shown in Fig. \ref{fig:sysmo}.  The precoded signals transmitted from the satellite towards the UTs can be grouped in a vector $\mathbf{x} = [x_1, ..., x_N]$ related to the received signals at the UTs $\mathbf{r} = [r_1, ..., r_K]$ by
\begin{equation}
\mathbf{r} = \mathbf{H}\mathbf{x} + \mathbf{z},
\label{eq:rx_signal}
\end{equation}
where $\mathbf{H}$ represents the $K \times N$ channel matrix and $\mathbf{z}$ contains the independent additive white Gaussian noise (AWGN) at the UTs modeled as circularly symmetric complex Gaussian random variables with zero mean and variance $\sigma^2$. 

The channel matrix elements are complex-valued and are denoted as $|h_{k_{n}}|{\rm e}^{\mathrm{j}\psi_{k_{n}}}$, with $|h_{k_{n}}|$ represents the attenuation and $\psi_{k_{n}}$  being the phase distortions  to the beam $n$ received by the $k$-th UT. To apply precoding technique, the intended modulated symbols $\textbf{s} = [s_1, ..., s_K]$ are multiplied by the $N \times K$ precoding matrix $\textbf{W}$, ($\textbf{H}\textbf{W} = \textbf{I}$) in such a way that \eqref{eq:rx_signal} becomes
\begin{equation}
\begin{split}
\mathbf{r}  &= \mathbf{H}\mathbf{W}\mathbf{s} + \mathbf{z}= \mathbf{s} + \mathbf{z}.
\end{split}
\label{eq:rx_signal_precoding}
\end{equation}

As such, the precoding matrix is calculated at the GW using the CSI from each UT. The CSI contains the channel estimation for each beam the $k$-th UT receives. Formally, it can be represented as $\hat{\mathbf{h}}_k = [\hat{h}_{k_{1}}, ..., \hat{h}_{k_{N}}]$, where $\hat{h}_{k_{n}}$ is the estimated complex channel coefficient. The CSI estimation\footnote{This operation is supported by the DVB-S2X standard that establishes the alternate transmission of precoded data and non-precoded pilots in its superframe (SF) format structure. Considering standard DVB-S2X SF-compliant terminals helps in estimating the real-time CSI to facilitate precoding at the GW. Thus, the GW can compensate for the differential frequency and phase distortions introduced by the conventional satellite transponder design, as discussed in our previous work \cite{DUNCAN2019ICSSC}.} is performed at the UT using the non-precoded pilots periodically sent by the GW. 

\section{Precoding Implementation}
\label{sec:precoding_implementation}
The DVB-S2X data streams are jointly precoded via the precoding block, which applies the precoding matrix $\mathbf{W}$ to the data symbols. According to \cite{ETSITR102376-2}, only certain fields of the DVB-S2X SF are precoded. For example, the precoding block is not activated at the start of SF (SOSF) and SF pilots. The SOSF is a known Walsh-Hadamard sequence that can be reliably detected at UTs even in high-interference conditions. On the other hand, the P-pilots are not precoded since they are used by the UTs to estimate the CSI, denoted by $\hat{\mathbf{H}}$, and to calculate the differential frequency and phase offset between the two spot beams. Note that the estimated CSI $\hat{\mathbf{H}}$ at the UTs is fed back through the return link and is repeatedly used to compute the precoding matrix coefficients by the GW.

\subsection{Calculation}
In our experiments, three precoding schemes are implemented at the GW, namely, two different types of MMSE precoding and optimal linear precoding. The simplest precoder implemented at the GW is the MMSE scheme in its basic form \cite{1391204}, which is given by
\begin{equation}\label{eq:mmse}
\mathbf{W}_{\textmd{MMSE}}= {\hat{\mathbf{H}}}^{\textmd{H}}(\hat{\mathbf{H}}{\hat{\mathbf{H}}}^{\textmd{H}}+\sigma^{2}\mathbf{I})^{-1},
\end{equation}
with $\sigma^2$ denoting the noise variance measured at the UT side.
Another type of precoding implemented at the GW is the so-called MMSE per-antenna power-constrained (PAC) in which the precoding matrix is calculated as
\begin{equation}\label{eq:mmsepac}
\mathbf{W}_{\textmd{MMSE-PAC}}= {\hat{\mathbf{H}}}^{\textmd{H}}(\hat{\mathbf{H}}{\hat{\mathbf{H}}}^{\textmd{H}}+\mathbf{\Lambda})^{-1},
\end{equation}
where the regularization $\mathbf{\Lambda}$ is a real diagonal matrix consisting of the Lagrangian dual variables. The optimal regularization $\mathbf{\Lambda}$ must satisfy
\begin{equation}\label{eq:mmsepaclamb}
\mathbf{\Lambda} \left(\textmd{diag}(\mathbf{W}\mathbf{W}^{\textmd{H}}) - \phi \mathbf{I} \right) = \mathbf{0},
\end{equation}
where $\phi$ is the available per-antenna transmit power. To find the optimal $\mathbf{\Lambda}$, we use a low-complexity iterative method as proposed in \cite{5585631} with its convergence proof. Note that, assuming symbols with unit average power, the squared Euclidean norms of the rows and columns of the precoding matrix respectively correspond to the per-antenna and per-beam power levels. The MMSE-PAC technique is particularly of interest as it yields a precoding matrix with all rows and columns normalized to have a squared norm of $\phi$. For example, by setting $\phi=1$, both per-antenna and per-beam power levels in $\mathbf{W}_{\textmd{MMSE-PAC}}$ are normalized to one. Nevertheless, further normalization steps are needed which may affect the optimality of the precoder.

The third precoding scheme is implemented based on the technique presented in \cite{opt_lin_prec}. We refer to this scheme as optimal linear (OPTL) precoding. Let $\mathbf{h}_i$ and $\mathbf{w}_i$ respectively denote the channel and the precoding vectors for the $i$-th UT such that $\mathbf{H} = [\mathbf{h}_1, \mathbf{h}_2,...,\mathbf{h}_K]^{\textmd{H}}$ and $\mathbf{W}_{\textmd{OPTL}}= [\mathbf{w}_1, \mathbf{w}_2,...,\mathbf{w}_K]$. The OPTL technique aims to find the optimum precoding vectors by solving the following virtual uplink problem as follows:
\begin{align}\label{eq:optl}
\underset{\mathbf{u}_i,p_i}{\min}& \quad \sum_{i=1}^K p_i\nonumber\\
\mathrm{s.t.}& \quad \frac{p_i\mathbf{u}_i^{\textmd{H}}\mathbf{h}_i\mathbf{h}_i^{\textmd{H}}\mathbf{u}_i}{\displaystyle\sum_{j\neq i} p_j\gamma_j \mathbf{u}_i^{\textmd{H}}\mathbf{h}_j\mathbf{h}_j^{\textmd{H}}\mathbf{u}_i+1} \geq 1, \quad i = 1,2,...,K\nonumber\\
&\quad \|\mathbf{u}_i\|^2 = 1, \quad i = 1,2,...,K,
\end{align}
where $\gamma_j$ is the given signal-to-noise-and-interference ratio (SNIR) requirement for the $j$-th UT. The optimal precoding vectors are then obtained as $\mathbf{h}_i = \sqrt{p_i}\mathbf{u}_i$, for all $i\in\{1,2,...,K\}$. To solve the problem in \eqref{eq:optl}, an iterative algorithm is presented in \cite{opt_lin_prec} which is composed of two main steps, namely, beamformer update and power control update. The corresponding algorithm is summarized as in Algorithm \ref{alg:optl}, wherein the superscript $t$ refers to the iteration number, with $Q$ being the total number of iterations.

\subsection{Complexity Analysis}
The complexity of designing precoding in (\ref{eq:mmsepac}) has two fold: (i) calculating $\mathbf{W}$; and (ii) computing Algorithm \ref{alg:optl}, which yields $\mathbf{\Lambda}$. In this context, the concrete takeaway messages of designing precoding in (\ref{eq:mmsepac}) are as follows:
\begin{itemize}
    \item In both (i) and (ii), only linear vector and matrix calculation operations are required, leading to low complex designing of precoding  $\mathbf{W}$.
    \item To calculate $\mathbf{\Lambda}$ via using Algorithm \ref{alg:optl} in (ii), it is not necessary to consider greedy iterative search so that the number of $Q$ is small. This will be justified later in analyzing the numerical results.
\end{itemize}
In the following, we resume the main operations needed to determine the precoding matrix using the proposed technique.
(i) Matrix summation or differentiation - $\mathcal{O}(K^{0})$\\
(ii) Matrix multiplication for $K\times K$ matrices $\mathbf{A}$ and $\mathbf{B}$ - $\mathcal{O}(K^{3})$\\
(iii) Matrix inversion - $\mathcal{O}(K^{3})$\\
(iv) Multiplication and division of two vectors a and b - $\mathcal{O}(K)$\\
(v) Scalar multiplied with vector or matrix -  $\mathcal{O}(K^{0})$.\\

\noindent The rest of the operation complexities can be neglected. 
Assuming $N_{\textmd{iter}}$ iterations and $L_{\textmd{frame}}$ the length of the frame to which a single precoder is applied, the following complexity can be found:
\begin{itemize}
    \item The complexity of calculating $\mathbf{W}$: $\mathcal{O}(3 K^{3})$ (over bundle frames)
    \item Complexity of calculating $\mathbf{\Lambda}$ into $\mathbf{W}$: $\mathcal{O}(6 K^{2} N_{\textmd{iter}})$ (over bundle frames)
    \item Complexity of application of precoding matrix $\mathbf{W}$: $\mathcal{O}(K^{2})$ (per symbol)
\end{itemize}

It is clear that the application of the precoders (both standard MMSE or the proposed one), taking into account a system of 16 beams, counts more than the computation itself since the ratio between the symbol rate and the Bundle rate is about 67920 for Format 2 and about 16984 for Format 3. The above analysis does not take into account operations that are or are not parallelizable within an FPGA. This complexity analysis is crucial to justify the exclusion of optimization-based techniques. According to the literature, the complexity of each iteration for solvers such as the interior point method used to solve Semi-Definite Programming problems in CVX is $\mathcal{O}(K^{6})$ \cite{christopoulos2014weighted}. Although low-complexity methods or optimized techniques could potentially reduce this complexity, their performance has not been validated in satellite systems. 

\subsection{Normalization}
To address the requirements of the GW feeder link, we have to ensure a unit output signal power for each transmit antenna. From the optimization problem in \eqref{eq:optl}, it follows that the OPTL design satisfies the sum power constraint but has no control over the per-antenna powers. Therefore, our implementation of the OPTL precoding includes an additional row-wise normalization so that each row of the precoding matrix $\mathbf{W}_{\textmd{OPTL}}$ has a unit Euclidean norm.

\begin{algorithm}[t!]
\caption{\textbf{: OPTL Precoding}}
\label{alg:optl}
1:  $\text{\bf{input:}}\; \{\mathbf{h}_i, \gamma_i\}_{i=1}^K$\\
2:  $\text{\bf{output:}}\; \{\mathbf{w}_i\}_{i=1}^K$\\
3: $t = 0,\, p_i^{(0)} = 1,\, i=1,2,...,K$\\
4: \textbf{while }{{\it the terminating condition is not met}}\textbf{ do }\\
5: \textit{Beamformer update:}\\
6: {$\mu_i^{(t+1)} \gets \underset{\|\mathbf{u}_i\|=1}{\max} \frac{{p_i}^{(t)}\mathbf{u}_i^{\textmd{H}}\mathbf{h}_i\mathbf{h}_i^{\textmd{H}}\mathbf{u}_i}{\displaystyle\sum_{j\neq i} {p_j}^{(t)}\gamma_j \mathbf{u}_i^{\textmd{H}}\mathbf{h}_j\mathbf{h}_j^{\textmd{H}}\mathbf{u}_i+1}$}\\
7: \textit{Power control update:}\\
8: $p_i^{(t+1)} \gets (\gamma_i/\mu_i)\, p_i^{(t)}$\\
9: $t \gets t+1$\\
10: \textbf{end while }\\
11: $\mathbf{q} = [\gamma_1\sigma^2,\gamma_2\sigma^2,...,\gamma_K\sigma^2]^{\textmd{T}}$\\
12:  $[\mathbf{F}]_{i,j} = \begin{cases} \mathbf{u}_i^{\textmd{H}}\mathbf{h}_i\mathbf{h}_i^{\textmd{H}}\mathbf{u}_i, & i=j, \\  -\gamma_i \mathbf{u}_j^{\textmd{H}}\mathbf{h}_j\mathbf{h}_j^{\textmd{H}}\mathbf{u}_j, & i\neq j\end{cases}$\\
13:  $\mathbf{p} = \mathbf{F}^{-1}\mathbf{q}$\\
14: $\mathbf{w}_i = \sqrt{[\mathbf{p}]_i} \, \mathbf{u}_i$.
\end{algorithm}

\section{Experimental validation}\label{sec:experimental}
\begin{figure}[t]
\centering
\includegraphics[height=2.3in, width=3in]{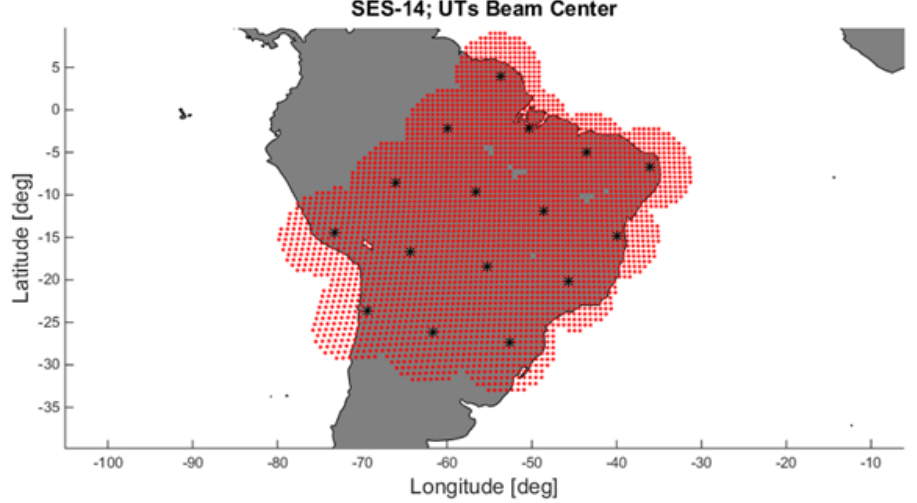}
\caption{SES-14 scenario SES14_A1, 16 Beams over South America, Users at Beam Center.}
\label{fig:user_location}\vspace{-0.2in}
\end{figure}

The experimental validation corresponding to the 16$\times$16 AWGN channel emulation is depicted in this section.
For the satellite link, we emulated the SES-14 satellite. The beam patterns, HPA, and thermal noise were simulated to emulate the SES-14 satellite link. SES-14 is operational since 2018.  The hybrid satellite provides C-and Ku-band wide beam coverage, as well as Ku-and Ka-band  HTS coverage across the Americas and the North Atlantic region. Also, SES-14 has two spot beams in the Ku-band covering a part of Western Europe and the United Kingdom. We show the approximated spot beams in Fig. \ref{fig:user_location}.

\begin{figure}[t!]
\centering
\includegraphics[height=3.7in, width=3in]{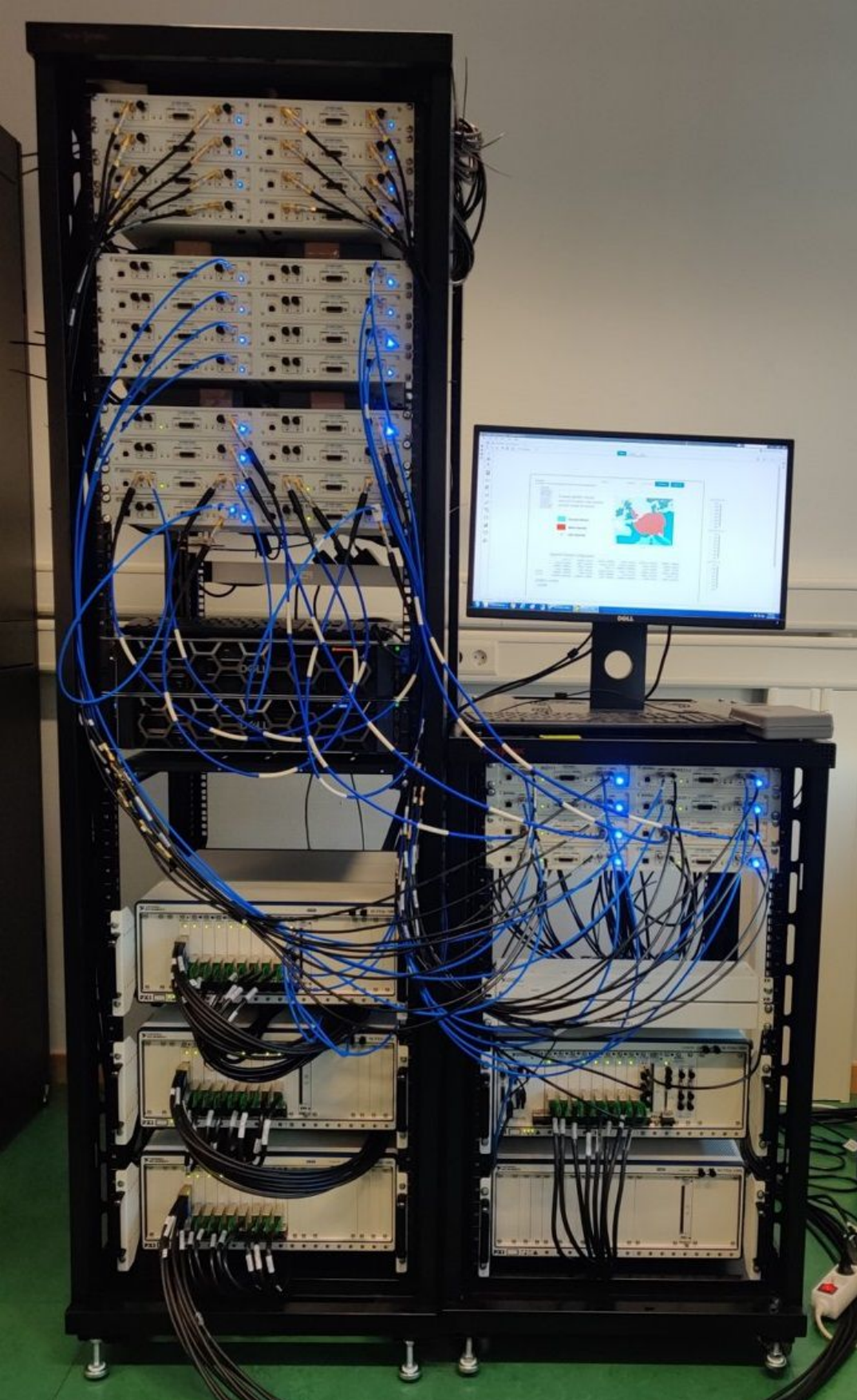}
\caption{In-Lab demo equipment.}
\label{fig:su1}
\end{figure}

\begin{figure}
\centering
\includegraphics[width=\linewidth]{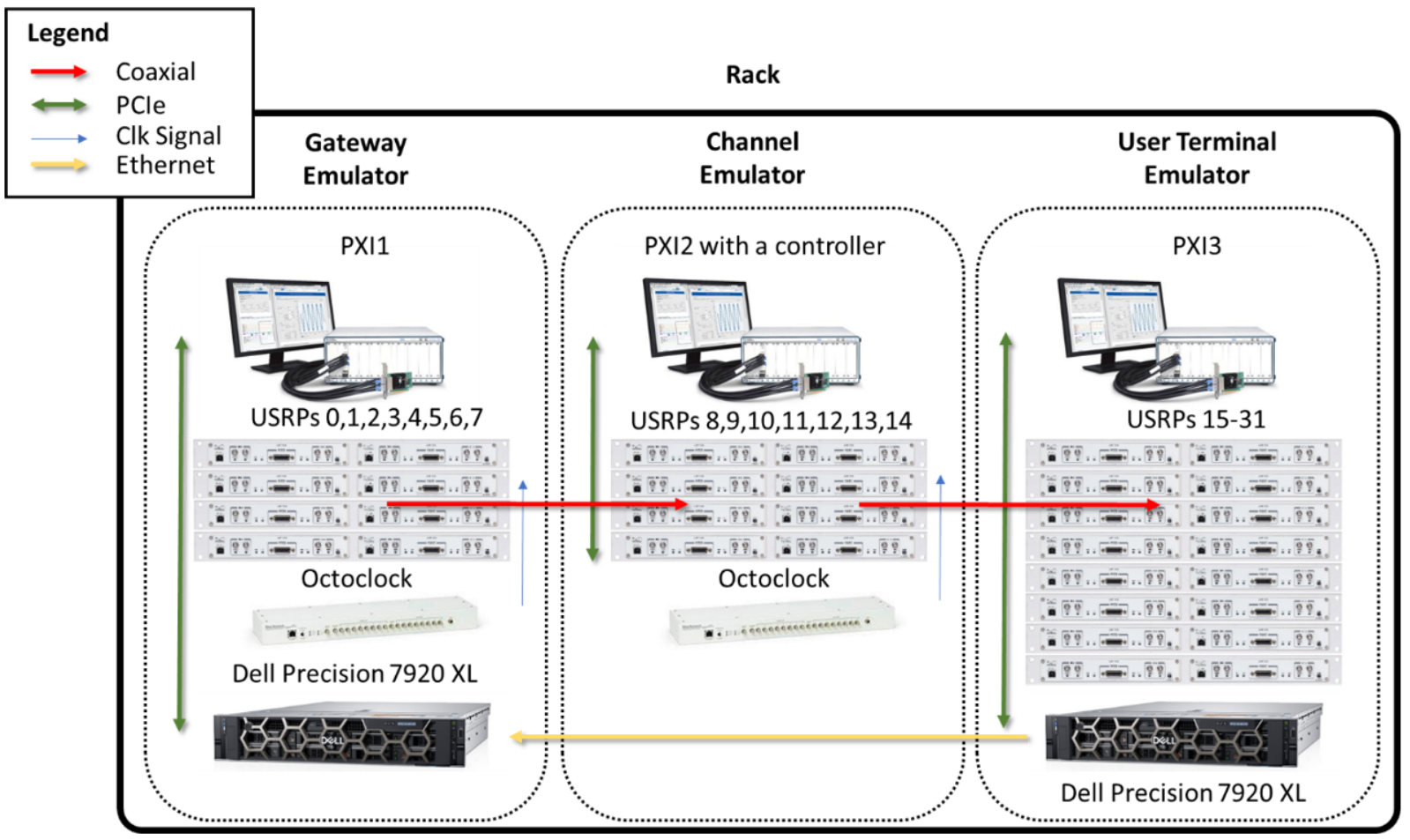}
\caption{In-Lab demo setup overview.}
\label{fig:su2}
\end{figure}

\vspace{0.15in}
The In-Lab equipment overview and setup are shown as in Fig. \ref{fig:su1} and Fig. \ref{fig:su2}, respectively. The in-house developed MIMO end-to-end satellite emulator includes several key components: a multichannel GW with precoding functionality, a MIMO satellite channel emulator (ChEm), multiple independent  UTs, and a return-link emulator.  

To summarize the system: the GW subsystem is responsible for producing data packets in accordance with the DVB-S2X standard, utilizing the SF format~II structure, and implementing the chosen precoding technique. The ChEm replicates the whole forward link chain, from the intermediate frequency (IF) input of the GW block up-converter (BUC), toward the low-noise block down-converter (LNB) IF output at the user terminal. It emulates the impairments present in the GW, the payload, the downlink channel, and the UTs. The UTs implement the synchronization and decoding features in the DVB-S2X-compliant receivers and perform the CSI estimation. Finally, the return-link emulator allows each UT to send its estimated CSI to the GW. Additional details on the demonstrator subsystems can be found in \cite{DUNCAN2019ICSSC}. The current implementation is an upgraded version of the setup presented in the cited work, extending the MIMO emulation capabilities from $6\times6$ to $16\times16$ channels.
It is also worth noting that the developed DVB-S2X modems performance has been successfully tested over a live GEO link~\cite{krivochiza2021end}.

The hardware components associated with the In-Lab demo are summarized in Table \ref{tab:lab}. In addition to this, two computer servers are employed to run the In-Lab test-bed demo, where each server has an Intel Xeon Gold 6148 CPU, 48 GB of RAM, and 1 TB of SSD storage. The servers run on the Microsoft Windows 10 Pro operating system and National Instruments (NI) LabVIEW NXG software. Additionally, each server is equipped with a peripheral component interconnect express (PCIe)-8381 host interface card to control the NI peripheral component interconnect (PCI) extensions for instrumentation (PXI) chasses from the servers. A PXI remote control module is placed in the system slot of the PXI chassis and a host interface card is used in the host server. This allows the host computer to establish a PCI Express connection to the chassis using a compatible multisystem extension interface (MXI)-Express cable and simultaneously operate all the universal software radio peripherals (USRPs).

The performance of the In-Lab test is evaluated for different  FFR configurations using MMSE precoder with  PAC, MMSE precoder with maximum power constraint (MPC), and 4FR. These comparisons for FFR are made with respect to three conditions of CSI. The first CSI condition is that precoders are calculated considering the ideal channel matrix, whereas the second and third CSI conditions are based on the calculation of precoders with CSI errors corresponding to the normal CSI and ESA NGW project. The parameters used for all experiments are summarized in Table \ref{tab:lab2}.

\begin{table}[ht]
\caption{Summary of the hardware components of the In-Lab demo.}
\centering
\begin{tabular}{|l|ccc|}
\hline
\multicolumn{1}{|c|}{\multirow{2}{*}{\textbf{Item}}} & \multicolumn{3}{c|}{\textbf{Number/Subsystem}} \\ \cline{2-4} 
\multicolumn{1}{|c|}{} & \multicolumn{1}{l|}{\textbf{GW}} & \multicolumn{1}{l|}{\textbf{UTs}} & \multicolumn{1}{l|}{\textbf{ChEm}} \\ \hline
\begin{tabular}[c]{@{}l@{}}USRP\\  RIO 2945R\end{tabular} & \multicolumn{1}{c|}{8} & \multicolumn{1}{c|}{16} & 8 \\ \hline
PXIe-1085 & \multicolumn{1}{c|}{1} & \multicolumn{1}{c|}{1} & 1 \\ \hline
FlexRIO   field programmable gate array (FPGA) & \multicolumn{1}{c|}{1} & \multicolumn{1}{c|}{0} & 1 \\ \hline
Octoclock & \multicolumn{3}{c|}{1 (common)} \\ \hline
\end{tabular}
  \label{tab:lab}
\end{table}

\begin{table}[ht]
\caption{System parameter used for the experiments.}
  \centering
\begin{tabular}{ | p{1.8in} | p{1.4in} |}
\hline
 \textbf{Parameter} & \textbf{Value} \\ 
 \hline
 Orbit & GEO, 47.5 deg west \\  
 \hline
 Coverage & 16 beams over South America \\
 \hline
  No. of active beams & 16  \\
 \hline
 Transponder BW & 54 MHz \\
 \hline
 Symbol rate & 20 MSPS \\
 \hline
 Roll-off factor & 0.2 \\
 \hline
 Symbol Rate & 20 MSPS \\
 \hline
Carrier frequency & 11.73 GHz \\
 \hline
 Air interface & DVB-S2X, SuperFraming \\
 \hline
 No. of carriers per transponder & 1 \\
 \hline
Adaptive coding and modulation (ACM) margin & 0.6 dB \\
 \hline
 Oversampling factor & 4 \\
 \hline
\end{tabular}
  \label{tab:lab2}
\end{table}
\vspace{-0.2in}
\section{Results and Discussion}
In this section, the performance of the precoders is evaluated with respect to the received precoded SNIR amongst beams versus the per-beam peak power. The results are as a function of the satellite per antenna power ($\textmd{P}_{\textmd{sat}}$) minus the output back-Off (OBO). It is noteworthy that the FFR configurations use half of the power of the 4FR since the throughput is calculated on the per-polarization SNIR and then multiplied by a factor of two (both polarizations). Specifically, with PAC we refer to the per-line normalization of the precoding matrix so that the transmission power is maximized for each antenna, whereas with MPC we refer to the power scaling applied to the whole precoder based on the maximum value calculated on per-line in the precoding matrix. In order to account for the channel amplifier (CAMP) effect, and in particular, for the normalization effect of the automatic level control (ALC) which normalizes over the signal power, a per-line re-scaling of the precoder is performed and termed as per-antenna power re-scaling (PAR). Hereby, a fair comparison of MMSE-PAC is performed with the MMSE-PAR.

\begin{figure}[t!]
\centering
\includegraphics[height=2.5in, width=3.5in]{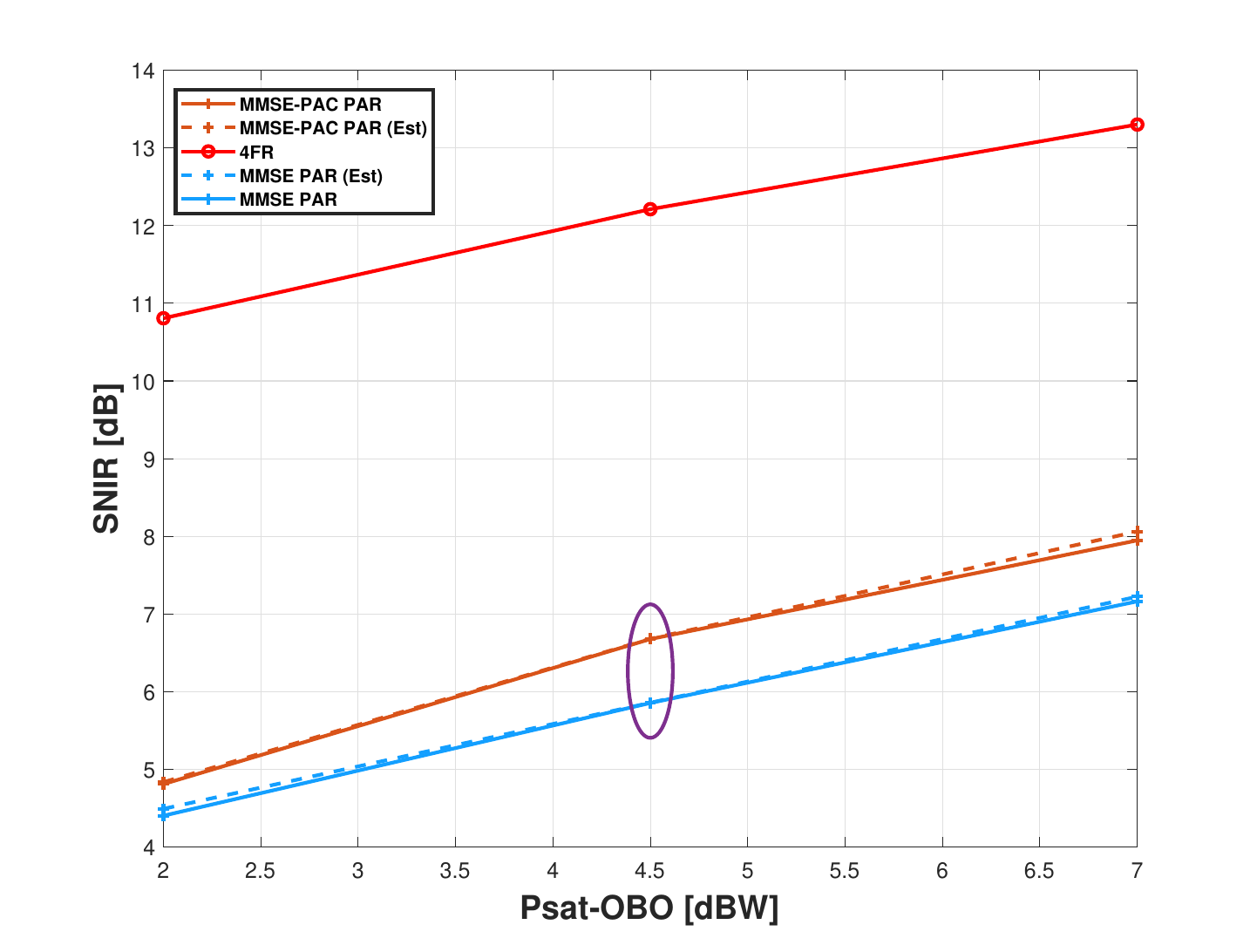}
\caption{Obtained Average SNIR as a function of Psat using CSI errors.}
\label{fig:pre0}
\end{figure}

\begin{figure}[t!]
\centering
\includegraphics[height=2.5in, width=3.5in]{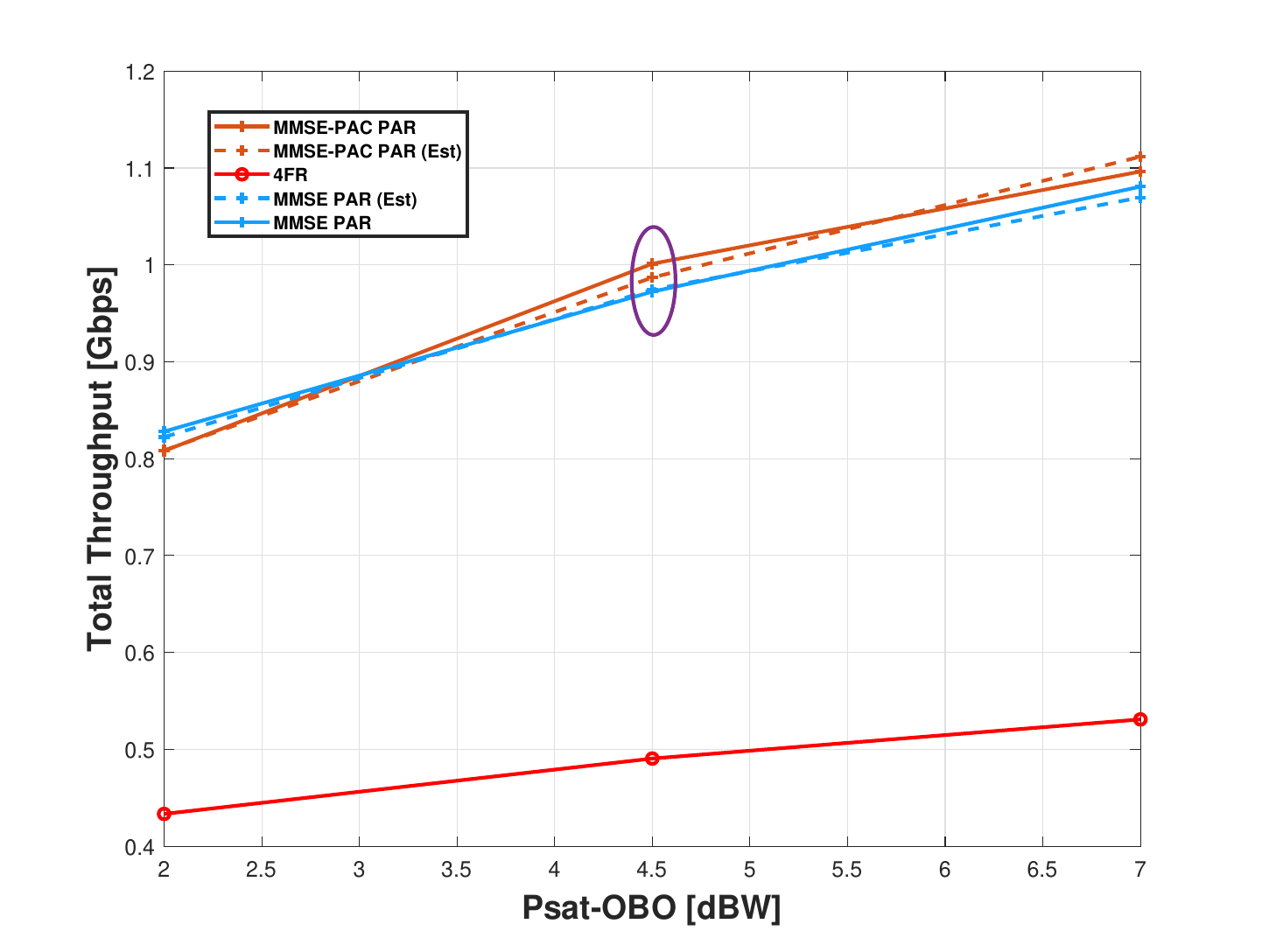}
\caption{ Obtained total throughput as a function of Psat using CSI errors.}
\label{fig:pre1}\vspace{-0.2in}
\end{figure}

\begin{figure}
\centering
\includegraphics[height=2.9in, width=3.5in]{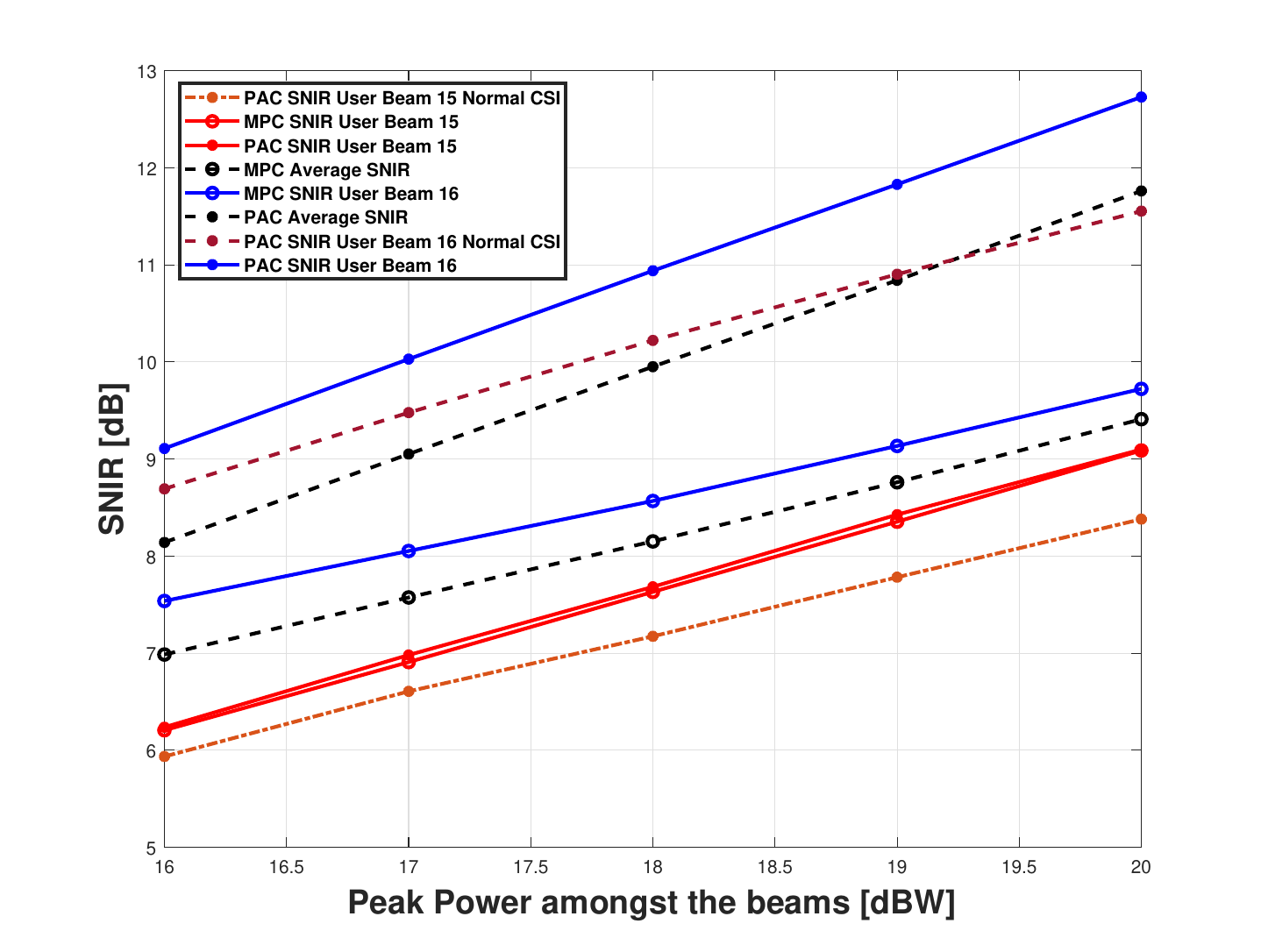}
\caption{Obtained SNIR for beams 16 and 15 as a function of Psat using CSI errors given by the new receiver
architecture.}
\label{fig:pre2}\vspace{-0.2in}
\end{figure}


From Fig. \ref{fig:pre0} and Fig. \ref{fig:pre1}, it can be visualized that for the operational point, which is 4.5 dBW, MMSE-PAC has better performance in terms of SNIR and total throughput as compared to MMSE-PAR. The average SNIR for MMSE-PAC at 4.5 dBW is 6.8 dB for estimated and actual ones, whereas, for MMSE-PAR, it is around 5.9 dB for estimated and actual.


While Fig. \ref{fig:pre1} aims at evaluating the average received SNIR performance over the whole coverage, in Fig. \ref{fig:pre2} the objective is to highlight the received SNIR performance versus ($\textmd{P}_{\textmd{sat}}$) by focusing on some specific users in some specific beams. The received SNIR performance is more sensitive to errors in the estimation (mainly nullification) for higher power transmissions. To highlight this aspect, the curve for PAC SNIR User Beam 16 has to be compared with the curve for PAC SNIR User Beam 16 Normal CSI, and the curve for PAC SNIR User Beam 15 has to be compared with the curve for PAC SNIR User Beam 15 Normal CSI. 

\begin{figure}
\centering
\includegraphics[height=2.9in, width=3.5in]{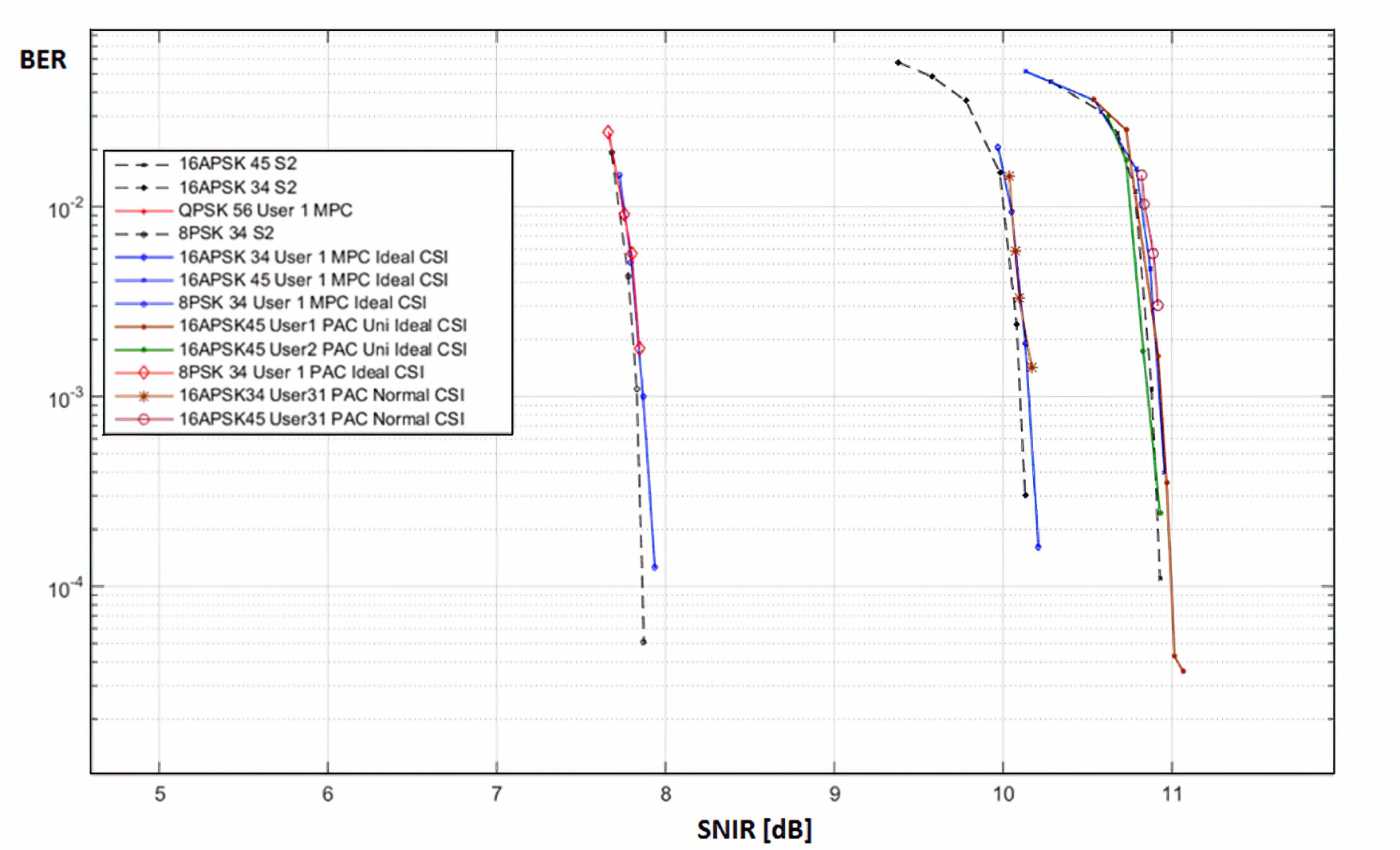}
\caption{BER performance as a function of SNIR for some beams and some MODCODs.}
\label{fig:pre3}
\end{figure}

In order to validate the analysis based on the SNIR behaviours, in Fig. \ref{fig:pre3}, end-to-end low-density parity code (LDPC) based simulations for bit error rate (BER) performance versus received precoded SNIR are shown. Three MODCODs (modulation constellation and code rate scheme) from DVB-S2X are used in order to test the technique at different SNIR levels. It is worth noting that the curves are here reported with a BER value down to $10^{-4}$ since the aim is not to test the error floor of the specific MODCOD but to test the behaviour of the LDPC decoding process when precoding is employed in the system.

Finally, based on Scenario SES14 A1 as illuminated earlier in Fig. \ref{fig:user_location}, 16$\times$16 scenario has been tested with the In-Lab demo with all the users located in the beam centers. Accordingly, in the case of MMSE precoding, we used two different power constraints: in the first case, we rely on the constant power spectral density (PSD) assumption between the FFR and 4FR systems, whereas in the second case, we rely on a constant total power constraint. In particular, we use as a reference the power of the precoding case, so the PSD of the precoding is not changing, while the PSD of the 4FR in this configuration increases, leading to an increase of the SNIR in the 4FR constant total power scenario. For the MMSE-PAC precoding technique, only the constant PSD power constraint has been tested. Corresponding to the above, the results are summarized from Table III to Table \ref{tab:lab6}, whereby one can note the significant advantages brought by the MMSE-PAC precoding technique in terms of average SNIR and system throughput. 

$$
\begin{aligned}
&\text { TABLE III: Scenario SES14_A1 MMSE SNIR Comparison (in dB) }\\
&\begin{array}{|l|l|l|l|}
\hline \textbf{ UT } & \begin{array}{l}
\textbf { FFR } \\
\textbf { (MMSE) }
\end{array} & \begin{array}{l}
\textbf { 4FR (const. } \\
\textbf { PSD) }
\end{array} & \begin{array}{l}
\textbf { 4FR (const. } \\
\textbf { total power) }
\end{array} \\
\hline \text { UT1 } & 11.77 & 15.96 & 20.45 \\
\hline \text { UT2 } & 12.99 & 17.23 & 21.42 \\
\hline \text { UT3 } & 12.54 & 15.56 & 19.85 \\
\hline \text { UT4 } & 12.14 & 16.49 & 20.75 \\
\hline \text { UT5 } & 13.61 & 17.06 & 21.59 \\
\hline \text { UT6 } & 14.95 & 17.58 & 22.08 \\
\hline \text { UT7 } & 12.16 & 15.22 & 19.87 \\
\hline \text { UT8 } & 12.46 & 16.58 & 21.07 \\
\hline \text { UT9 } & 13.57 & 14.67 & 19.24 \\
\hline \text { UT10 } & 12.73 & 15.55 & 19.78 \\
\hline \text { UT11 } & 11.95 & 15.65 & 20.05 \\
\hline \text { UT12 } & 12.21 & 15.41 & 19.61 \\
\hline \text { UT13 } & 13.61 & 16.43 & 20.87 \\
\hline \text { UT14 } & 13.34 & 16.04 & 20.18 \\
\hline \text { UT15 } & 12.14 & 15.70 & 19.64 \\
\hline \text { UT16 } & 10.51 & 15 & 19.28 \\
\hline \hline \textbf { Average } & 12.67 & 16.07 & 20.36 \\
\hline
\end{array}
\end{aligned}
$$

$$
\begin{aligned}
&\text { TABLE IV: Scenario SES14_A1 MMSE Precoding Throughput}\\
&\text {Comparison (in Mbps) }\\
&\begin{array}{|l|l|l|l|}
\hline \textbf{ UT } & \begin{array}{l}
\textbf { FFR } \\
\textbf { (MMSE) }
\end{array} & \begin{array}{l}
\textbf { 4FR (const. } \\
\textbf { PSD) }
\end{array} & \begin{array}{l}
\textbf { 4FR (const. } \\
\textbf { total power) }
\end{array} \\
		\hline \text {UT1} & 16.11 & 5.36 & 6.69 \\ 
        \hline \text {UT2} & 17.14 & 6.01 & 6.69 \\ 
		\hline \text {UT3} & 17.14 & 5.36 & 6.69 \\ 
		\hline \text {UT4} & 16.11 & 5.36 & 6.69 \\ 
		\hline \text {UT5} & 17.14 & 6.01 & 6.69 \\ 
		\hline \text {UT6} & 21.45 & 6.01 & 6.69 \\ 
        \hline \text {UT7} & 16.11 & 5.36 & 6.69 \\ 
        \hline \text {UT8} & 16.11 & 5.36 & 6.69 \\ 
		\hline \text {UT9} & 17.14 & 5.36 & 6.69 \\ 
		\hline \text {UT10} & 17.14 & 5.36 & 6.69 \\ 
		\hline \text {UT11} & 16.11 & 5.36 & 6.69 \\ 
		\hline \text {UT12} & 16.11 & 5.36 & 6.69 \\ 
        \hline \text {UT13} & 17.14 & 5.36 & 6.69 \\ 
        \hline \text {UT14} & 17.14 & 5.36 & 6.69 \\ 
		\hline \text {UT15} & 16.11 & 5.36 & 6.69 \\ 
		\hline \text {UT16} & 14.48 & 5.36 & 6.69 \\ 
        \hline \hline \textbf{System} & 268.70 & 87.77 & 107.10 \\ 
\hline
\end{array}
\end{aligned}
$$

\addtocounter{table}{2}
\begin{table}[t]\caption{Scenario SES14_A1 MMSE-PAC SNIR Comparison (in dB)}
	\label{tab:lab5}
	\centering
	\begin{tabular}{|c|c|c|}
		\hline 
		\textbf{UT} & \textbf{FFR (MMSE-PAC) } &  \textbf{4FR (const. PSD)}\\ \hline
		UT1 & 10.95 & 15.96   \\ \hline
        UT2 & 13.20  & 17.23   \\ \hline
		UT3 & 12.10  & 15.56  \\ \hline
		UT4 & 11.72  & 16.49   \\ \hline
		UT5 & 12.85  & 17.06   \\ \hline
		UT6 & 15.18  & 17.58  \\ \hline
        UT7 & 11.35  & 15.22   \\ \hline
        UT8 & 11.86  & 16.58   \\ \hline
		UT9 & 16.11  & 5.36 \\ \hline
		UT10 & 17.14 & 6.01 \\ \hline
		UT11 & 17.14 & 5.36 \\ \hline
		UT12 & 16.11 & 5.36  \\ \hline
        UT13 & 17.14 & 6.01 \\ \hline
        UT14 & 21.45  & 6.01  \\ \hline
		UT15 & 16.11 & 5.36 \\ \hline
		UT16 & 16.11 & 5.36  \\ \hline \hline
        \textbf{Average} & 14.78 & 11.03 \\ \hline
	   \end{tabular}\\
\end{table}

\begin{table}[t]\caption{Scenario SES14_A1 MMSEPAC Throughput (in Mbps)}
	\label{tab:lab6}
	\centering
	\begin{tabular}{|c|c|c|}
		\hline 
		\textbf{UT} & \textbf{FFR (MMSE-PAC) } &  \textbf{4FR (const. PSD)}\\ \hline
		UT1 & 17.9696 & 6.7122    \\ \hline
        UT2 & 26.8489   & 6.7122    \\ \hline
		UT3 & 17.9696   & 6.7122  \\ \hline
		UT4 & 17.9696   & 6.7122    \\ \hline
		UT5 & 26.8489   & 6.7122    \\ \hline
		UT6 & 26.8489   & 6.7122  \\ \hline
        UT7 & 17.9696   & 6.7122    \\ \hline
        UT8 & 17.9696   & 6.7122    \\ \hline
		UT9 & 17.14   & 5.36 \\ \hline
		UT10 & 17.14 & 6.01 \\ \hline
		UT11 & 16.11 & 5.36 \\ \hline
		UT12 & 16.11 & 5.36  \\ \hline
        UT13 & 17.14 & 5.36 \\ \hline
        UT14 & 17.11  & 5.36  \\ \hline
		UT15 & 16.11 & 5.36 \\ \hline
		UT16 & 14.48 & 5.36  \\ \hline \hline
        \textbf{System} & 301.72 & 97.22 \\ \hline
	   \end{tabular}\\
\end{table}

\section{Conclusion}
\label{sec:conclusions}


In this work, we implemented optimal linear precoders at the GWs for the multibeam GEO satellite to mitigate the issue of multiple-user interference. We further demonstrated standard DVB-S2X SF-compliant terminals capable of estimating real-time CSI to facilitate precoding at the GW. Using the SF structure, the GW can compensate for the differential frequency and phase introduced by the conventional satellite transponder design.  In the conducted field test, we demonstrated end-to-end SNIR and coded throughput gains of precoded communications over the actual satellite forward link. The results show that terminal-specific data can be transmitted to independent UTs through the same physical channel by utilizing closed-loop precoding over a multi-beam satellite. Further, we demonstrated that precoding techniques enabled with FFR communications in SATCOM outperform conventional 4FR schemes. Moreover, performance evaluations reveal that MMSE-PAC outperforms MMSE-PAR in SNIR and throughput at 4.5 dBW operational point. Notably, the analysis highlights the scheme's sensitivity to CSI errors at higher power transmissions, with end-to-end LDPC-based simulations confirming significant SNIR and throughput improvements with MMSE-PAC. This advancement not only enhances data transmission efficiency but also paves the way for more robust and adaptable satellite communication systems, highlighting the potential for future implementations in various satellite communication applications.

\vspace{-4mm}
\bibliographystyle{IEEEtran}
\bibliography{References}
\end{document}